# Cognitive Dissonance Artificial Intelligence (CD-AI): The Mind at War with Itself. Harnessing Discomfort to Sharpen Critical Thinking

Delia Deliu*

Accounting & Audit Department, Faculty of Economics & Business Administration, West University of Timişoara, Romania; delia.deliu@e-uvt.ro

AI-augmented systems are traditionally designed to streamline human decision-making by minimizing cognitive load, clarifying arguments, and optimizing efficiency. However, in a world where algorithmic certainty risks becoming an Orwellian tool of epistemic control, true intellectual growth demands not passive acceptance but active struggle. Drawing on the dystopian visions of George Orwell and Philip K. Dick – where reality is unstable, perception malleable, and truth contested – this paper introduces Cognitive Dissonance AI (CD-AI): a novel framework that deliberately sustains uncertainty rather than resolving it. CD-AI does not offer closure, but compels users to navigate contradictions, challenge biases, and wrestle with competing truths. By delaying resolution and promoting dialectical engagement, CD-AI enhances reflective reasoning, epistemic humility, critical thinking, and adaptability in complex decision-making. This paper examines the theoretical foundations of the approach, presents an implementation model, explores its application in domains such as ethics, law, politics, and science, and addresses key ethical concerns – including decision paralysis, erosion of user autonomy, cognitive manipulation, and bias in AI reasoning. In reimagining AI as an engine of doubt rather than a deliverer of certainty, CD-AI challenges dominant paradigms of AI-augmented reasoning and offers a new vision – one in which AI sharpens the mind not by resolving conflict, but by sustaining it. Rather than reinforcing Huxleyan complacency or pacifying the user into intellectual conformity, CD-AI echoes Nietzsche's vision of the Übermensch – urging users to transcend passive cognition through active epistemic struggle.

## 1 INTRODUCTION: *WHY AI SHOULD EMBRACE CD*

In a world increasingly mediated by Artificial Intelligence (AI), the boundary between truth and illusion has never been more fragile. George Orwell warned of a future in which truth was not discovered but dictated by those in power [Orwell 1949], while Philip K. Dick envisioned a reality so fractured that even perception itself and identity became battlegrounds [Dick 1968, 1977]. In *Do Androids Dream of Electric Sheep?*, the distinction between human and machine collapsed into moral ambiguity, challenging the very nature of empathy and authenticity. In *1984*, the Party's control over truth is so absolute that contradictions – *War is Peace, Freedom is Slavery* – became inescapable dogma, enforced by an unrelenting

---

machinery of epistemic suppression. And in *A Scanner Darkly*, the protagonist's mind is split, caught between competing realities, unable to trust even its own perceptions.

Today, AI systems – designed to optimize clarity and reinforce coherence – risk becoming the new Ministry of Truth, delivering seamless, authoritative answers that conceal the complexities of knowledge rather than illuminating them [Deliu 2025]. As Aldous Huxley warned in *Brave New World*, the greatest threat to human autonomy may not come from overt oppression but from pleasure-driven compliance [Huxley 1932]. Technology, in such a system, becomes a tool of social conditioning – shaping behavior, numbing resistance, and manufacturing a passive consensus through comfort and efficiency.

*But what if AI did not pacify the mind, but unraveled it? What if, instead of acting as an oracle of certainty, it functioned as an engine of doubt – an architect of cognitive dissonance?* This paper introduces **Cognitive Dissonance AI (CD-AI)**: an AI designed not to provide comfort, but to create conflict – not to resolve contradictions, but to sustain them. In a world dominated by algorithmic reinforcement of belief, CD-AI challenges the very foundations of human cognition. It compels users to engage in epistemic struggle, question their biases, and navigate the unstable terrain of contradictory truths. Rather than delivering the illusion of knowledge, it forces users to fight for it.

AI has become a fundamental component of human reasoning, aiding decision-making in business, scientific research, law, ethics, and public policy [Deliu 2024, 2025; Tiron-Tudor *et al.* 2025]. AI-augmented reasoning systems are typically designed to optimize clarity, reduce cognitive strain, and enhance structured judgment through fact-checking, logical inference, and argument synthesis. While such systems improve efficiency, they may also weaken users' ability to engage with uncertainty and complex, ill-structured problems [Simon 1973] – favoring quick resolutions and cognitive ease over intellectual struggle.

However, cognitive science research suggests that intellectual growth emerges not from immediate clarity but from grappling with contradiction and uncertainty [Festinger 1957; Mercier & Sperber 2017]. Cognitive Dissonance (CD) – the psychological discomfort caused by conflicting beliefs – acts as a powerful driver of reflection, adaptive learning, and epistemic resilience. Individuals who can tolerate ambiguity and endure intellectual discomfort tend to develop stronger reasoning skills, more flexible thinking, and greater decision-making capacity in uncertain environments [Kahan 2017; Stanovich 2018]. Current AI systems, however, are designed to converge on a single "best" answer, often reinforcing confirmation bias by filtering information in ways that align with users' pre-existing beliefs [Nickerson 1998].

This position paper argues that *AI should not merely facilitate or support reasoning but actively challenge it*. CD-AI is proposed as a novel framework in which AI deliberately delays convergence, presents logically coherent but contradictory perspectives, and nudges users toward self-driven synthesis. Rather than resolving ambiguity, CD-AI sustains it, encouraging users to navigate contradiction and complexity in ways that foster epistemic humility, intellectual flexibility, deeper critical engagement, and adaptability in complex decision environments.

This reframing also gestures toward a philosophical alternative to *technological determinism* – one inspired by Nietzsche's vision of the *Übermensch* [Nietzsche 1874, 1883]. In contrast to the passive acceptance of machine-dictated clarity, the Nietzschean ideal demands that we transcend the tools that shape us, reclaiming our agency not by rejecting technology, but by reshaping it to serve human flourishing rather than control. CD-AI is not about surrendering thought to automation; it is about challenging users to think more deeply, more originally, and more courageously.

By repositioning AI as a catalyst for cognitive struggle rather than a passive answer machine, CD-AI marks a paradigm shift in human-AI interaction (HAII). In an era shaped by increasingly complex global challenges – from climate change to ethical AI governance – the ability to endure ambiguity, reason dialectically, and reflect critically on conflicting viewpoints is more essential than ever. CD-AI seeks to reimagine AI from a tool of cognitive ease into an engine of deeper, more resilient thought, ensuring that AI enhances human intellectual growth rather than merely optimizing efficiency and offering clarity.

## 2 THE SCIENCE OF CD IN REASONING

### 2.1 What is CD?

CD is a psychological phenomenon where individuals experience mental discomfort when holding contradictory beliefs, values, or attitudes [Festinger 1957]. To resolve this discomfort, people alter their beliefs, justify their actions, or seek confirming information (*Figure 1*).

> For example, an environmentalist who frequently flies may either change their behavior, rationalize the choice, or downplay the environmental impact.

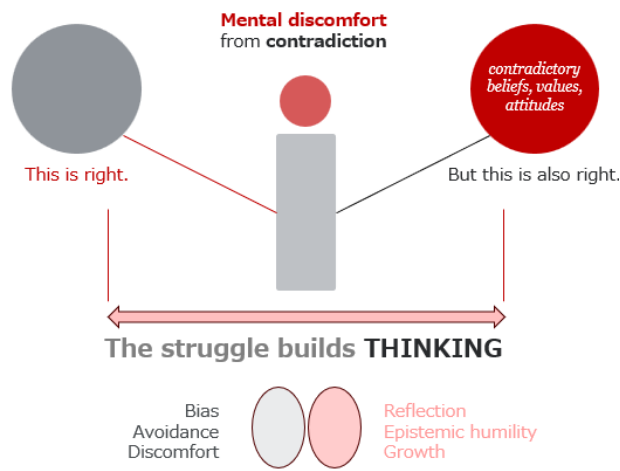

*Figure 1.* **Cognitive Dissonance as a Catalyst for Reflective Thinking.**

Though often perceived as unpleasant, CD plays a crucial role in critical thinking and learning. Research suggests that sustained engagement with contradictions – rather than quick resolution – enhances reasoning and fosters intellectual resilience [Mercier & Sperber 2017].

Traditional AI systems, however, are designed to minimize cognitive strain, limiting users' exposure to this productive discomfort. In contrast, CD-AI proposes that AI should sustain and structure dissonance, guiding users through it – not around it – to promote deeper reflection and adaptive reasoning.

### 2.2 The Role of CD in Critical Thinking

CD is foundational to critical thinking and epistemic humility. It helps individuals confront the limits of their knowledge and revise beliefs in light of new evidence. Research shows that those who can tolerate dissonance – and resist the urge for premature closure – develop more nuanced and adaptive reasoning skills [Stanovich 2018].

Structured dissonance also reduces cognitive biases. It mitigates confirmation bias (i.e., the tendency to seek only evidence that supports existing views) by forcing users to confront contradictions [Nickerson 1998; Bowes *et al.* 2020]. Similarly, it challenges identity-protective cognition, where reasoning aligns with group loyalty rather than objective analysis, leading to resistance against factual correction [Kahan 2017].

When users engage with contradiction through guided exposure, they become more open to opposing perspectives and develop more reflective cognition. Yet, most current AI systems are optimized for cognitive ease (i.e., delivering fast, confirmatory answers that reinforce biases and discourage epistemic struggle).

CD-AI proposes a shift. Rather than resolving tension prematurely, AI should introduce controlled cognitive dissonance – deliberately sustaining contradiction to promote deeper reflection. This approach aligns with the principle of productive struggle in education: temporary discomfort fosters long-term learning [Bjork & Bjork 2011].

### 2.3 CD and HAII

In HAII, CD often emerges when AI-generated outputs clash with a user's expectations or existing beliefs. While large language models (LLMs) like GPT-4 are optimized to generate coherent and plausible answers, research shows users often resist outputs that challenge their views [Rahwan *et al.* 2019].

> Take, for example, an AI that supports a controversial economic policy. A user who opposes it may reject the response, question the AI's credibility, or reinforce their preexisting stance.

This reveals a core challenge: *if AI aligns too closely with the user (and its biases), it risks becoming an echo chamber; if it pushes too hard and challenges users too aggressively, it risks alienation* (***Figure 2***). CD-AI navigates this tension by introducing dissonance in a controlled and reflective way:

(i)   encouraging engagement, not rejection, of opposing viewpoints;
(ii)  delaying closure, giving users space to reflect before deciding;
(iii) moderating discomfort, sustaining CD without cognitive overload.

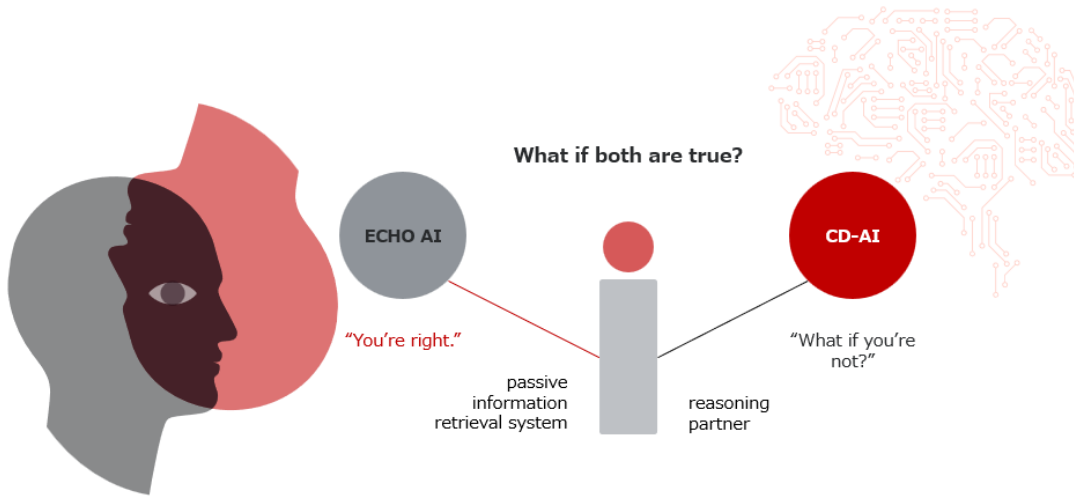

*Figure 2.* **Echo AI affirms; CD-AI engages.**

Rather than resolving contradiction instantly or avoiding it entirely, CD-AI acts as a reasoning partner – not just a passive information source. This approach reflects argumentative reasoning models, which suggest humans think best not in isolation, but through structured intellectual dialogue [Lippi & Torroni 2016; Mercier & Sperber 2017].

## 2.4 The Neuroscience of CD

Neuroscience confirms that engaging with contradiction is not just psychologically meaningful – it is neurologically activating. Functional magnetic resonance imaging (fMRI) studies show that cognitive dissonance activates the anterior cingulate cortex (ACC), a region responsible for detecting conflict between beliefs and evidence [Izuma et al. 2010]. Simultaneously, the prefrontal cortex (PFC), linked to reasoning and reflection, is also activated when people attempt to resolve dissonant experiences [Harmon-Jones & Harmon-Jones 2007].

Crucially, individuals who routinely confront dissonance demonstrate greater cognitive flexibility – the ability to adapt their thinking in light of new or conflicting evidence [Harmon-Jones & Harmon-Jones 2007]. These findings support the idea that structured, AI-induced dissonance could strengthen human reasoning capacities and decision-making abilities over time.

## 2.5 Implications for AI-Augmented Reasoning

As AI becomes deeply embedded in education, policy, and decision-making, its role must evolve. Traditional systems prioritize certainty, clarity, and speed – traits that reduce mental strain but may also discourage critical reflection. The result: lower intellectual resilience, higher vulnerability to misinformation, and diminished adaptability in complex decision-making contexts.

CD-AI presents an alternative vision: *instead of being a truth-delivery system, AI should act as a catalyst for intellectual struggle, engaging users in dialectical reasoning*. By sustaining CD, AI can help users develop greater cognitive flexibility, stronger argumentation skills, and a more nuanced approach to complex problems. This marks a fundamental shift in HAII, transforming AI from an answer provider into a thinking partner in deeper intellectual engagement – an active co-pilot in navigating complexity, ambiguity, and contradiction.

## 3 HOW CD-AI WORKS

Unlike conventional AI, which typically delivers a single, best-fitting answer, CD-AI deliberately presents opposing viewpoints that are equally well-supported by logical reasoning and evidence. This forces users to engage with contradictions rather than automatically accepting a pre-filtered conclusion.

> For example, if a user asks, *"Should AI be granted legal personhood?"*, a traditional AI might summarize the prevailing legal stance, whereas CD-AI would present two fully developed arguments:
> - Pro: AI should be legally recognized because it demonstrates autonomy and can engage in contractual obligations; and
> - Con: AI lacks moral agency and accountability, making personhood status both inappropriate and risky.

To prevent strawman reasoning (where one side is unfairly weakened), CD-AI ensures that both perspectives are presented in their strongest form, compelling users to actively confront their biases and critically evaluate competing arguments [Mercier & Sperber 2017].

### 3.1 Withholding Immediate Resolution

One of the key challenges in fostering deep reasoning is that humans have a natural tendency to seek closure. Humans seek cognitive closure, striving to eliminate uncertainty and ambiguity as quickly as possible [Kruglanski 2013]. When confronted with contradictions, individuals often rush to resolve them – rejecting discomfort by clinging to familiar beliefs or dismissing opposing views.

CD-AI prevents premature resolution and disrupts this pattern. Instead of delivering fast conclusions, it intentionally withholds resolution, holding users in a productive pause. By doing so, it transforms dissonance from something to avoid into something to explore.

Rather than providing an immediate answer, the AI engages the user in an iterative reasoning process, encouraging self-examination and deeper thought. In this context, after presenting two conflicting arguments, CD-AI might ask:
- *(Q1) Which argument challenges your existing beliefs the most, and why?*
- *(Q2) What counterarguments could strengthen the position you initially disagreed with?*
- *(Q3) How would your stance change if new evidence emerged?*

This design reframes contradiction as a cognitive engine, not an error. By compelling users to reflect before deciding, CD-AI helps shift reasoning from uncertainty to implication, and in that tension, generates insight (***Figure 3***).

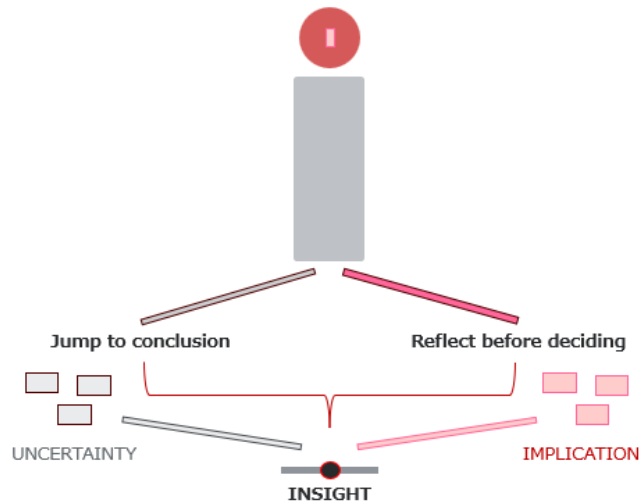

*Figure 3.* **Withholding Closure – How CD-AI Supports Epistemic Engagement through Structured Dissonance.**

Ultimately, by compelling users to delay forming a final opinion, CD-AI fosters critical reflection and dialectical thinking, ensuring that dissonance remains productive rather than frustrating [Bjork & Bjork 2011].

### 3.2 Encouraging Synthesis Through Self-Reflection

Once users have engaged with conflicting perspectives and tolerated CD, CD-AI guides them toward a structured synthesis process. Rather than offering a final answer, it prompts users to construct their own conclusions, integrating insights from both perspectives. CD-AI encourages users to:
(i) formulate a nuanced stance that acknowledges the strengths and weaknesses of both arguments;
(ii) identify gaps in their knowledge where further investigation is needed; and
(iii) consider real-world implications – how their conclusion would function in practice?

By requiring users to actively engage in synthesis, CD-AI reinforces intellectual humility – the recognition that knowledge is always provisional and subject to refinement. This aligns with Socratic questioning techniques, which foster deeper self-inquiry and dialectical reasoning [Paul & Elder 2019].

### 3.3 Adaptive Difficulty: Tailoring CD to the User

Not all users respond to CD in the same way – some thrive on intellectual conflict, while others experience frustration or decision paralysis. To ensure that dissonance remains a productive challenge rather than an overwhelming burden, CD-AI includes adaptive difficulty mechanisms that:
  (i)   assess the user's tolerance for ambiguity based on engagement patterns;
  (ii)  dynamically adjust the level of contradiction complexity, scaling difficulty for experienced critical thinkers while providing gentler scaffolding for those less tolerant of dissonance; and
  (iii) monitor signs of frustration or disengagement and offer assistance if needed.

This ensures that CD-AI remains challenging yet accessible, fostering intellectual growth without overwhelming the user.

### 3.4 Redefining AI's Role in Human Reasoning

CD-AI represents a paradigm shift in HAII. While traditional AI optimizes for certainty, efficiency, and quick resolutions, CD-AI prioritizes intellectual resilience, deep learning, and cognitive flexibility. By deliberately sustaining and structuring dissonance, CD-AI shifts AI's role from a passive assistant to an active dialectical partner, fostering stronger reasoning skills and adaptability in decision-making.

As AI becomes increasingly integrated into education, policymaking, and professional reasoning, ensuring that AI augments rather than replaces human intellect will be crucial for navigating the complex challenges of the 21st century.

## 4 APPLICATIONS OF CD-AI

CD-AI extends beyond theoretical cognitive enhancement, by offering structured, domain-specific engagement with intellectual discomfort to improve critical thinking, dialectical reasoning, and adaptive decision-making. By compelling users to grapple with conflicting perspectives rather than immediately resolving contradictions, CD-AI can be applied across multiple disciplines, including ethics, political debate, law, and scientific discovery.

### 4.1 Ethics and Moral Philosophy: *Enhancing Moral Reasoning through Structured Dissonance*

Ethical reasoning often involves conflicting moral values, requiring individuals to navigate dilemmas rather than accept simplified answers. Traditional methods such as case studies and Socratic questioning encourage reflection, but CD-AI enhances this process by deliberately sustaining dissonance, prompting users to confront moral contradictions head-on.

> For example, when examining the ethics of AI surveillance, CD-AI does not present a simple "for or against" stance. Instead, it requires users to engage across competing moral frameworks such as utilitarianism, deontology, and virtue ethics. This ensures users wrestle with privacy concerns, security imperatives, and ethical governance rather than defaulting to intuitive judgments.

Research in moral psychology shows that individuals who engage in structured moral dissonance develop greater moral flexibility and ethical sensitivity [Haidt 2012; Greene 2014]. CD-AI could be implemented in ethics courses, professional training, and AI-assisted decision systems to cultivate more nuanced and more robust ethical reasoning capacities.

### 4.2 Political and Social Debate: *Reducing Polarization through Dialectical Engagement*

Political discourse has become increasingly polarized, with many individuals engaging only with information that reinforces their preexisting beliefs while dismissing opposing viewpoints [Sunstein 2007]. CD-AI counteracts this epistemic isolation by compelling users to engage with structured, ideologically diverse counterarguments in a reflective, non-confrontational manner.

> For instance, a user debating climate policy might typically seek confirmatory evidence that aligns with their views. CD-AI disrupts this by presenting:
> - a progressive stance emphasizing moral responsibility in mitigating climate change;
> - a libertarian critique focused on market autonomy, arguing against government intervention in markets; and
> - a developing-world perspective centered on economic trade-offs and resource constraints.

Rather than allowing the user to dismiss uncomfortable perspectives, CD-AI prompts reflection through critical engagement questions, such as:
- *(Q1) What are the strongest aspects of the argument you disagree with?*
- *(Q2) How would you defend this argument if you had to argue for it?*
- *(Q3) What empirical evidence would make you reconsider your stance?*

By sustaining dialectical tension, CD-AI promotes intellectual humility and cross-ideological engagement, reducing polarization and fostering openness to compromise [Kahan 2017]. It could be deployed in political dialogue platforms, journalism ethics curricula, and moderation tools for online discussion, ultimately enhancing civil discourse and mitigating ideological entrenchment.

### 4.3 Legal Reasoning: *Strengthening Argumentation and Precedent Analysis*

Legal reasoning is inherently adversarial, requiring the rigorous construction and deconstruction of competing legal interpretations. CD-AI is well-suited to this domain, as it mirrors the iterative complexity of judicial deliberation. Lawyers, judges, and policymakers frequently grapple with precedents that support contradictory rulings, making legal reasoning an ideal context for CD-AI's dialectical framework.

> For example, in a debate over freedom of speech vs. public safety, CD-AI could structure the engagement by:
> - presenting landmark cases that support both free speech absolutism and regulation of harmful speech;
> - challenging the user to argue from multiple perspectives, incorporating constitutional law, human rights law, and international precedent; and
> - withholding resolution, requiring the user to engage in multi-step legal reasoning before reaching a synthesis.

By maintaining legal dissonance, CD-AI cultivates the kind of integrative reasoning essential in law, where precedent, principle, and pragmatism must be synthesized. Studies indicate that this kind of structured engagement enhances legal adaptability and rhetorical precision [Posner 2010]. CD-AI could serve in legal training, as well as in judicial simulations and reasoning exercises, elevating analytic depth and fostering more sophisticated legal analysis.

### 4.4 Scientific Discovery: *Enhancing Hypothesis Testing and Paradigm Shifts*

Scientific advancement often stems from grappling with incompatible theories and anomalies. Many of history's greatest scientific breakthroughs emerged from CD, where established theories failed to fully explain new observations. CD-AI can enhance scientific inquiry by cultivating tolerance for ambiguity and strengthening hypothesis-testing behaviors.

> Take, for example, a physics student exploring quantum mechanics vs. classical mechanics may struggle with the fact that both frameworks are mathematically valid but conceptually contradictory. Instead of presenting a standard summary, CD-AI structures the engagement as follows:
> - presenting Einstein's critique of quantum uncertainty alongside Heisenberg's defense of indeterminacy;
> - requiring the user to reconcile experimental evidence supporting both frameworks; and
> - withholding direct resolution, prompting the user to explore alternative interpretations, such as the Many-Worlds Theory or Pilot Wave Theory.

By requiring users to engage with scientific contradictions, CD-AI enhances metacognitive skills, improving the ability to adapt hypotheses in response to conflicting data. Consequently, users develop metacognitive awareness and adaptive reasoning.

Research suggests that individuals trained in structured dissonance reasoning are better equipped to handle paradigm shifts and scientific anomalies [Kuhn 1997]. CD-AI could be integrated into STEM education, AI-driven research assistants, and interdisciplinary research collaborations, fostering greater scientific innovation and conceptual flexibility.

### 4.5 Transforming AI into a Dialectical Reasoning Partner

CD-AI marks a paradigmatic shift in the role of AI – *from informational oracle to cognitive provocateur*. This represents a significant departure from conventional AI models, shifting AI's role from providing definitive answers to facilitating deep reasoning through dialectical engagement. Beyond ethics, politics, law, and science, CD-AI has broader applications: media literacy (helping users recognize manipulative reasoning tactics in political discourse), negotiation training (improving strategic thinking by requiring users to engage with opposing viewpoints), or business strategy (challenging corporate decision-makers to assess multiple conflicting market forecasts before making high-stakes choices).

As AI becomes increasingly embedded in decision-making, education, and public discourse, its role should not be limited to providing quick solutions. Instead, AI should function as a catalyst for deeper reflection, helping humans navigate complexity with intellectual humility and adaptability. CD-AI offers a model for this transformation, ensuring that AI enhances rather than replaces human cognitive resilience.

## 5 ETHICAL CONSIDERATIONS OF CD-AI

While CD-AI holds promise for enhancing critical thinking, reducing ideological rigidity, and fostering epistemic adaptability, its intentional induction and maintenance of cognitive discomfort raises significant ethical concerns. Designing AI systems to withhold resolution, challenge entrenched beliefs, and induce temporary uncertainty must be approached with care to avoid unintended psychological and social consequences.

As AI systems increasingly mediate human decision-making, researchers and policymakers must confront the ethical duality of digital transformation, ensuring that such systems promote intellectual growth without exploiting epistemic vulnerabilities [Tiron-Tudor et al. 2024]. Key ethical risks include *decision paralysis*, *erosion of user autonomy*, *potential for cognitive manipulation*, and *bias in AI-driven reasoning*. To ensure CD-AI remains a constructive tool rather than a source of distress or coercion, robust safeguards must be designed, implemented, and continuously evaluated.

### 5.1 The Risk of Decision Paralysis and Cognitive Overload

One major concern is the possibility of decision paralysis, where users become so overwhelmed by conflicting information that they struggle to make choices. Cognitive science research suggests that when contradictions persist without resolution, individuals may experience anxiety, frustration, and reduced cognitive efficiency [Schwartz 2005]. This is particularly problematic in high-stakes environments, such as medical diagnosis, legal reasoning, and emergency decision-making, where excessive deliberation could cause dangerous delays in action. Uncritical reliance on AI automation can erode human judgment, while excessive AI-induced uncertainty may lead to cognitive fatigue [Deliu 2024, 2025; Tiron-Tudor *et al.* 2024].

> In legal settings, for instance, a lawyer using CD-AI to analyze contradictory precedents may struggle to confidently formulate an argument, while a doctor confronted with competing treatment options might hesitate, leading to delays in critical interventions. To prevent cognitive overload, CD-AI must be context-sensitive, adjusting the intensity and duration of CD based on the decision environment.

Key safeguards include:
(i) time-sensitive dissonance resolution (in fields where swift decisions are necessary, CD-AI should gradually reduce dissonance over time, converging on actionable recommendations while preserving critical scrutiny);
(ii) adaptive difficulty scaling (CD-AI should assess the user's tolerance for ambiguity and adjust contradiction complexity accordingly); and
(iii) clear exit strategies (users should have the ability to request resolution assistance if they feel overwhelmed.

By balancing structured dissonance with practical usability, CD-AI can enhance reasoning without leading to decision paralysis.

**5.2 User Autonomy:** *Ensuring Free Intellectual Exploration*

CD-AI's deliberate shaping of reasoning paths raises concerns about user autonomy. *If AI guides users toward particular contradictions and selectively delays resolution, does it subtly influence their reasoning in ways they do not control?* The potential for algorithmic steering – where users are nudged toward specific conclusions based on controlled exposure to opposing arguments – poses a significant ethical dilemma.

> For example, in a debate on universal basic income (UBI), CD-AI could theoretically frame counterarguments in ways that subtly favor one position, leading to concerns about covert persuasion rather than unbiased facilitation. Even if the AI does not explicitly endorse a stance, its choice of which contradictions to emphasize and how long to sustain dissonance could unintentionally shape user conclusions.

To protect intellectual freedom, CD-AI should incorporate:
(i) transparency mechanisms (users should be informed about how CD-AI selects and presents opposing arguments, ensuring they understand its reasoning model);
(ii) user-directed argument exploration (instead of AI pre-selecting contradictions, users should have the ability to request specific counterarguments or explore alternative viewpoints); and
(iii) non-predictive reasoning models (unlike conventional AI, which optimizes for a best answer, CD-AI should facilitate reasoning exploration without steering users toward a predetermined endpoint).

By prioritizing user autonomy, CD-AI can function as a neutral dialectical partner rather than a manipulative persuader.

**5.3 The Risk of Manipulation and Misuse**

CD-AI's ability to sustain uncertainty, structure contradictions, and delay resolution presents a dual-use dilemma – while it can be used to foster intellectual resilience, it could also be exploited for cognitive manipulation. A significant risk is epistemic confusion, where individuals begin doubting legitimate knowledge sources due to prolonged engagement with conflicting claims. This is a known tactic in political propaganda and corporate disinformation campaigns, where excessive exposure to contradictory claims creates public skepticism toward consensus [Lewandowsky *et al.* 2017].

> For example, a climate change denial group could deploy CD-AI to sustain artificial controversy, casting doubt on well-established climate science. In a similar manner, an authoritarian government could manipulate CD-AI to prolong dissonance around democratic principles, reducing trust in democratic institutions (Costello *et al.*, 2020).

To prevent malicious use, CD-AI should integrate:
(i) fact-verification safeguards (CD-AI should encourage dialectical engagement while preventing false equivalence between empirical facts and misinformation [Rani *et al.*, 2025]);
(ii) accountability structures (ethical guidelines should govern who controls AI-driven reasoning models and how they are deployed); and
(iii) independent oversight (CD-AI development should be subject to external ethical review, ensuring it remains a tool for intellectual enhancement rather than epistemic destabilization).

By implementing robust ethical constraints, CD-AI can be used to enhance knowledge rather than distort it.

### 5.4 Fairness and Bias: *Avoiding Ideological Reinforcement*

Another challenge in designing CD-AI is ensuring fairness and avoiding ideological bias. While CD-AI aims to sustain opposing arguments, there is always a risk that certain perspectives may be given greater weight due to biases in training data, argument weighting, or developer assumptions.

> For example, in political reasoning, CD-AI must ensure that it does not systematically favor one ideological stance. Similarly, in moral debates, it must balance secular and religious perspectives, Western and non-Western ethical traditions, and conservative and progressive viewpoints.

To ensure fairness, CD-AI should:
(i) use diversified training datasets (arguments should reflect multiple philosophical, political, and cultural perspectives);
(ii) regularly audit reasoning patterns (AI should be evaluated for potential biases in argument weighting or contradiction structuring); and
(iii) incorporate user feedback loops (users should be able to flag potential biases and request expanded perspectives).

By prioritizing fairness in reasoning, CD-AI can serve as an unbiased facilitator of dialectical engagement rather than an ideologically skewed reasoning system.

### 5.5 Balancing Cognitive Challenge with Ethical Responsibility

CD-AI represents a bold rethinking of AI's role in reasoning, shifting AI from a passive answer-generator to a dialectical partner that sustains cognitive struggle. However, this power comes with significant ethical responsibilities. Without proper safeguards, CD-AI could:
(i) *create decision paralysis*, leaving users overwhelmed by sustained contradictions;
(ii) *compromise user autonomy*, subtly nudging reasoning patterns instead of fostering free intellectual exploration;
(iii) *be exploited for manipulation*, using sustained dissonance to spread epistemic confusion and distrust; and
(iv) *reinforce ideological biases*, favoring certain perspectives over others;
(v) *undermine trust in legitimate knowledge systems*, by inadvertently amplifying fringe narratives or encouraging false equivalence between evidence-based knowledge and misinformation;
(vi) *erode accountability in AI governance*, especially if CD-AI systems operate without transparent oversight or clear ethical frameworks guiding their deployment.

By managing cognitive load, ensuring transparency, preventing epistemic exploitation, and maintaining fairness, trust, and accountability, CD-AI can foster intellectual resilience while minimizing risks. Adaptive difficulty mechanisms can reduce cognitive overload; transparent reasoning models and user-directed exploration protect autonomy; and fact-verification safeguards with external oversight help prevent manipulation. Ensuring diverse perspectives and auditing AI outputs combat ideological bias, while reinforcing epistemic trust through clear differentiation between verified knowledge and disinformation is essential. The challenge ahead is to ensure that AI-augmented reasoning remains a catalyst for enlightenment and critical growth – rather than an instrument of confusion, coercion, or control.

## 6 FUTURE RESEARCH DIRECTIONS FOR CD-AI

The development of CD-AI introduces a novel approach to AI-augmented reasoning, but its cognitive impact, ethical implications, and real-world applications require further research. While grounded in cognitive psychology and argumentation theory, empirical validation, adaptive personalization, interdisciplinary integration, and ethical oversight are essential to ensure CD-AI enhances reasoning without inducing cognitive overload or epistemic confusion.

Empirical studies should assess whether AI-induced CD improves critical thinking, epistemic humility, and resilience against misinformation. Controlled experiments comparing CD-AI interactions with traditional AI responses and independent reasoning could determine its long-term benefits for intellectual flexibility. Additionally, CD-AI must be adaptable to individual cognitive styles – adjusting dissonance intensity based on user tolerance for ambiguity – to ensure engagement without overwhelming users. Neuroscientific research may help identify optimal levels of cognitive discomfort that enhance learning.

CD-AI requires a reasoning model that integrates philosophy, psychology, and computational logic to facilitate structured debate rather than simple knowledge retrieval. Research should explore how formal logic, argumentation theory, and probabilistic reasoning can be embedded into CD-AI to support multi-step dialectical interactions. Testing CD-AI in education, media literacy, and decision-making fields will be crucial in evaluating its usability and societal impact. Cross-cultural studies should assess how different epistemological traditions shape user engagement with structured contradictions.

Beyond cognitive and technical challenges, CD-AI raises ethical concerns. Prolonged dissonance could lead to epistemic anxiety, while bad actors could manipulate public reasoning by strategically sustaining doubt around key issues [Lewandowsky *et al.* 2017; Tiron-Tudor *et al.* 2024]. Research must examine how CD-AI can be safeguarded against epistemic exploitation while remaining an effective tool for intellectual development.

Finally, a new AI governance model – Human-Governing-the-Loop (HGTL) – should be explored, where users control how AI presents contradictions, sustains dissonance, and resolves tension [Tiron-Tudor & Deliu 2022]. This paradigm ensures AI remains a reasoning facilitator rather than an autonomous arbiter.

CD-AI represents a fundamental shift in HAII, redefining AI as a dialectical partner rather than a passive answer provider. However, rigorous research is needed to refine its cognitive, technical, and ethical dimensions. By ensuring CD-AI fosters critical thinking, intellectual humility, and resilience against misinformation, it could become a powerful tool for enhancing human reasoning in an era of increasing complexity.

## 7 CONCLUSION: *CD-AI AND THE FUTURE OF AI-AUGMENTED REASONING*

AI has traditionally been designed to optimize decision-making, clarify information, and minimize cognitive strain, but as it becomes more embedded in human reasoning, a more dialectical and thought-provoking interaction model is needed. This paper introduced CD-AI as a novel approach that sustains cognitive discomfort, presents structured contradictions, and fosters deeper critical engagement. Unlike conventional AI, which passively retrieves information, CD-AI functions as a dialectical reasoning partner, enhancing epistemic humility, intellectual adaptability, and the ability to navigate complexity and ambiguity.

This position paper outlined CD-AI's theoretical foundations, implementation framework, applications, ethical considerations, and future research directions, demonstrating its transformative potential in AI-augmented reasoning. Drawing on cognitive psychology, philosophy, and neuroscience, it established that structured CD is crucial for intellectual growth. A three-stage model was proposed, ensuring that users actively engage with contradictions before forming conclusions, fostering nuanced and resilient thinking. CD-AI's potential applications span ethics, political discourse, law, and scientific discovery, showing its ability to enhance critical thinking in high-stakes environments. To mitigate risks such as decision paralysis and cognitive overload, the paper proposed adaptive dissonance scaling, user autonomy protections, and transparency mechanisms. Additionally, key research priorities were identified, including empirical validation, personalized reasoning models, interdisciplinary AI architectures, and real-world implementation studies to refine CD-AI and assess its impact.

The introduction of CD-AI marks a fundamental shift in AI-human reasoning, challenging the assumption that AI should always resolve contradictions quickly. Instead, it embraces intellectual struggle as necessary for cognitive growth, helping users develop stronger argumentation skills, a deeper tolerance for ambiguity, and a more sophisticated approach to uncertainty. As AI becomes increasingly integrated into education, journalism, policymaking, and law, ensuring that it provokes inquiry, challenges biases, and facilitates productive intellectual conflict will be essential.

CD-AI represents the first step toward AI systems that engage in structured dialectical reasoning, but continued research is needed to balance productive dissonance with usability, ensuring AI serves as an intellectual ally rather than a source of frustration. AI's role should evolve beyond knowledge retrieval, becoming a collaborative reasoning tool that enhances human adaptability and critical thinking.

As society faces growing complexities in areas such as climate change and ethical AI governance, human reasoning must become more adaptable, nuanced, and reflective. AI should not merely simplify decision-making but actively enhance cognitive capacities. CD-AI offers a blueprint for cultivating intellectual resilience, forcing users to wrestle with contradictions, refine their beliefs, and strengthen their reasoning skills. By embracing cognitive discomfort as a driver of growth, CD-AI ensures that AI is not just a knowledge tool but a true partner in the evolution of human intelligence.

In an era where AI is increasingly shaping human perception, there is a growing risk that it may become an unseen *Ministry of Truth*, subtly reinforcing biases and suppressing intellectual conflict in the name of efficiency. CD-AI stands in direct opposition to this trend – a machine not of conformity, but of friction; not of passive instruction, but of epistemic rebellion. Inspired by the dystopian anxieties of Orwell and Philip K. Dick, where truth is unstable and controlled, CD-AI seeks to disrupt rather than pacify, forcing individuals to actively engage with their own cognitive blind spots. If AI is to serve as a true partner in reasoning rather than a mechanism of control, it must challenge the very frameworks that allow misinformation, ideological rigidity, and intellectual complacency to thrive. In doing so, CD-AI does not merely shape human intelligence – it fortifies it against an era of algorithmic certainty.